\documentstyle[twocolumn,graphics,prl,aps]{revtex}
\begin{document}

\wideabs{
\draft
\title{
Quasiparticle transport in the vortex state of YBa$_2$Cu$_3$O$_{6.9}$
}

\author{ May Chiao, R. W. Hill$^\dagger$, Christian Lupien$^\dagger$,
Bojana Popi\'c$^\dagger$, Robert Gagnon and Louis
Taillefer$^\dagger$\\ }

\address{Canadian Institute for Advanced Research and\\
Department of Physics, McGill University,
Montr\'eal, Qu\'ebec, Canada H3A 2T8}

\date{\today}

\maketitle

\begin{abstract}

The effect of vortices on quasiparticle transport in cuprate
superconductors was investigated by measuring the low temperature
thermal conductivity of YBa$_2$Cu$_3$O$_{6.9}$ in magnetic fields up
to 8~T.  The residual linear term (as $T \to 0$) is found to increase
with field, directly reflecting the occupation of extended
quasiparticle states. A study for different Zn impurity concentrations
reveals a good agreement with recent calculations for a $d$-wave
superconductor.  It also provides a quantitative measure
of the gap near the nodes.
\end{abstract}

\pacs{PACS numbers: 74.70.Tx, 74.25.Fy} }

In unconventional superconductors, where the gap structure has nodes
along certain directions, it was argued by Volovik \cite{Volovik} that
the dominant effect of a magnetic field should be the Doppler shift of
extended quasiparticle states due to the presence of a superfluid flow
around each vortex.  Reports of a $\sqrt{H}$ field-dependence in the
heat capacity of YBa$_2$Cu$_3$O$_{6.9}$ (YBCO) have generally been
accepted as strong evidence for this effect \cite{Moler}.  However, a
similar field dependence is also observed in an $s$-wave superconductor
such as NbSe$_2$ \cite{Sonier}, where it can only arise from localized
states bound to the vortex core.

In this respect, it is interesting to look at heat conduction, to
which only delocalized states contribute.  So far, in all measurements
performed above 5~K or so, the thermal conductivity $\kappa$ is found
to drop with field \cite{Palstra,Cohn,KK}, much as it does in a clean
type-II superconductor like Nb \cite{Niobium}, and eventually levels
off to a roughly constant plateau \cite{Ong}.  In
Bi$_2$Sr$_2$CaCu$_2$O$_8$ (BSCCO), the crossover from drop to plateau
occurs abruptly, at a field whose magnitude increases with temperature
\cite{KK}.  Franz has recently proposed that the plateau is due to a
compensation between the increasing occupation of extended
quasiparticle states {\it \`a la Volovik} and a parallel increase in
the scattering of quasiparticles by vortices \cite{Franz}.  While this
may well be part of the explanation, it can hardly serve as a direct
confirmation of the ``Volovik effect''.  Neither can it explain the
abruptness of the onset of the plateau.  The situation remains quite
puzzling, in part because of the large phonon background at these
temperatures.

In this Letter, we present a study of heat transport which provides a
solid experimental basis for a description of quasiparticle properties
in terms of the field-induced Doppler shift of a $d$-wave spectrum due
to the superflow around vortices.  Furthermore, the good agreement
found with recent calculations by K\"ubert and Hirschfeld
\cite{Kubert2} allows us to conclude that in YBCO at low temperature,
impurity scattering is close to the unitarity limit and vortex
scattering is weak.  Moreover, it provides a measure of the two
parameters that govern the Dirac-like spectrum near the nodes,
responsible for all low-energy properties.

We have measured the thermal conductivity of three optimally-doped
untwinned single crystals of YBa$_2$(Cu$_{1-x}$Zn$_x$)$_3$O$_{6.9}$
down to 0.07~K, with a current along the $a$-axis of the basal plane
and a magnetic field along the $c$-axis.  The nominal Zn
concentrations are $x = 0$ (pure), 0.006 and 0.03, corresponding to
$T_c$~=~93.6, 89.2, and 74.6~K, respectively.  The data were taken by
sweeping the temperature at fixed field, and the field was changed at
low temperature (below 1~K).  Data taken by field-cooling the samples
gave the same results.  Given that the field of full penetration is
estimated to be about 1~T \cite{Brandt}, the lowest field point was
set at 2~T.  The largest source of absolute error is the uncertainty
in the geometric factor (approximately $\pm$ 10\%); however, the
relative error between different field runs on the same crystal is
only due to the fitting procedure used to extract the electronic
contribution (see below), which is less than 10\%.

Fig.~1 shows the measured thermal conductivity divided by temperature,
$\kappa/T$, as a function of $T^2$ in fields up to 8 T.  As discussed
in Ref.\cite{Taillefer}, the only reliable way to separate the
electronic contribution from the phonon background is to reach the
$T^3$ regime for the phonon conductivity (below about 130~mK), from
which a linear extrapolation to zero yields the purely electronic
term at $T$=0, labelled $\kappa_0/T$.

{\it Zero field.}--- Let us first focus on the results at $H$=0. A
linear term is observed at $T$=0, of comparable magnitude in all three
crystals.  From measurements on a number of pure (optimally-doped)
crystals of YBCO, we obtain an average value for ${\bf J}~||~{\bf a}$
\cite{explain}:
\begin{equation}
\frac{\kappa_{0}}{T} = 0.14\pm 0.03~{\rm mW~K^{-2}~cm^{-1}}.
\end{equation}
This residual conduction is associated with an impurity band whose
width $\gamma$ is a new energy scale relevant to all low-energy
properties in superconductors with gap zeros \cite{Graf}.  $\gamma$
grows with the impurity scattering rate $\Gamma$, in a way which
depends strongly on whether impurities act
\begin{center}
\begin{figure}
\resizebox{7.5cm}{!}{\includegraphics{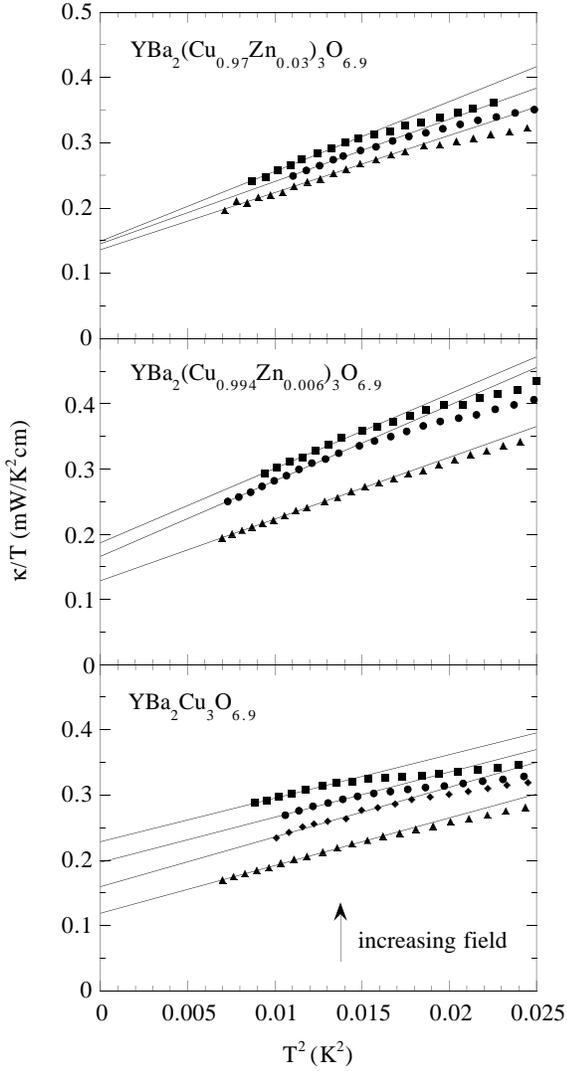}}
\vspace{0.5em}
\caption{Thermal conductivity divided by temperature vs $T^2$ of
YBa$_2$(Cu$_{1-x}$Zn$_{x}$)$_3$O$_{6.9}$ in a magnetic field, for H 
= 0 (triangles), 2 (diamonds), 4 (circles), and 8 T (squares).  The
bottom panel corresponds to $x$=0; middle, $x$=0.006; top, $x$=0.03.
The lines are linear fits to the data below 130~mK.}
\label{fig1}
\end{figure}  
\end{center}  
as Born or resonant scatterers.  In the unitarity limit, $\gamma$ is
maximum and given by $\gamma \simeq 0.61\sqrt{\hbar\Gamma\Delta_0}$,
where $\Delta_0$ is the gap maximum \cite{Kubert}.  This novel fluid,
deep in the superconducting state, should carry heat and charge.  Lee
was the first to point out the {\it universal} character of that
conduction as $T~\to$~0 (and frequency $\omega~\to~0$), for certain
gap topologies \cite{Lee}.  For the $d_{x^2-y^2}$ gap, the conductance
of a single CuO$_2$ plane is predicted to be $\sigma_{00} =
\frac{e^2}{2\pi\hbar}~\frac{2}{\pi}~\frac{v_F}{v_2}$, where $v_F$ and
$v_2$ are the two parameters governing the Dirac-like spectrum near a
node: $E(k) = \sqrt{\varepsilon_k^2 + \Delta_k^2} = \hbar
\sqrt{v_F^2k_1^2 + v_2^2k_2^2}$.  Here $(k_1,k_2)$ defines a
coordinate system whose origin is at the node, with $k_1$ normal to
the Fermi surface and $k_2$ tangential; thus $v_F$ is the Fermi
velocity and $v_2$ is the slope of the gap, at the node. The
conductivity is independent of impurity concentration as a consequence
of the exact compensation between the decreasing scattering time $\tau
\propto 1/\gamma$ and the growing residual quasiparticle density of
states, $N(0)\propto\gamma/( v_Fv_2)$, when impurities are added,
i.e. $\sigma_{00} \propto N(0) v_F^2 \tau \propto v_F/v_2$.

Graf and co-workers showed that the residual normal fluid should obey
the Wiedemann-Franz law at $T$=0 \cite{Graf}, so that the thermal
conductivity as $T\to$~0 in a $d$-wave superconductor should also be
universal, and given by:
\begin{equation}
\frac{\kappa_{00}}{T} = L_0~ \sigma_{00} = \frac{k_B^2}{3 \hbar}~
\frac{v_F}{v_2}~n,
\end{equation}
where $n$ is the number of CuO$_2$ planes per meter stacked along
the $c$-axis and $L_0 = \frac{\pi^2}{3} (\frac{k_B}{e})^2$.

In recent measurements of $\kappa(T)$ in YBCO, the universal limit was
observed \cite{Taillefer}.  It is also evident from the $H$=0 curves
in Fig.~1, noting that the scattering rate $\Gamma$ in the 3\% Zn
crystal is some 20 to 40 times larger than it is in the pure crystal.
Indeed, estimates from a combination of resistivity, microwave and
infrared measurements give $\hbar \Gamma/k_B T_{c0}$ = 0.014, 0.13,
and 0.54 for $x$ = 0, 0.006 and 0.03, respectively \cite{Taillefer}.
Note also that the impurity bandwidth in the 3\% Zn sample is a
sizable fraction of the gap maximum, so that corrections to the
universal limit are expected. Calculations by Sun and Maki give a 30\%
increase in $\kappa_0/T$ for 20\% $T_c$ suppression \cite{Sun95}, in
good agreement with the observed slight increase (see Fig. 2).

For the bi-layer structure of YBCO, $n=2/c$ where $c=11.7$ \AA~is the
$c$-axis lattice constant.  Combining this with Eqs. (1) and (2)
yields
\begin{equation}
\frac{v_F}{v_2} = 14 \pm 3.
\end{equation}
Taking $v_F = 1.2 \times$10$^7$~cm/s (see Ref.\cite{LeeWen}), one gets
$v_2 \simeq 1 \times$10$^6$~cm/s, which corresponds to a slope of the
gap at the node $d\Delta(\phi)/d\phi = \hbar k_F v_2 \simeq 380$~K,
very close to that estimated using the simplest $d$-wave gap,
$\Delta_0$cos$2\phi$, in the weak-coupling limit: $2 \Delta_0 = 2
\times 2.14 k_B T_c = 400$~K.  The same ratio $v_F/v_2$ governs the
linear drop in superfluid density with temperature when $\gamma\ll k_B
T\ll \Delta_0$.  In terms of penetration depth data, where $\lambda(T)
= \lambda(0) + \delta\lambda(T)$, one has $\delta\lambda(T) =
(\frac{c}{\omega_p}) 2\,{\rm ln}\,2 (\frac{k_BT}{\hbar k_Fv_2})$
\cite{LeeWen}, where $\omega_p^2 = 2\pi^2k_Fv_2\sigma_{00}$, so that
$\delta\lambda/T = 4.8$~\AA/K for the above values of $v_F$ and $v_2$.
This is in excellent agreement with the microwave measurements of
Zhang and co-workers, who obtain $\delta\lambda_a/T = 4.7$~\AA/K
\cite{Zhang}.

{\it Vortex state.}--- As seen in Fig.~1, the effect of a magnetic
field applied normal to the CuO$_2$ planes is to {\em increase}
$\kappa_a$ in YBCO.  To ensure that only the electronic contribution
is considered, the $T\to$~0 limit of $\kappa/T$ is plotted as a
function of field in Fig.~2, for each of the three crystals.  The
enhancement observed as $T\to$~0 is direct evidence for itinerant
quasiparticle excitations {\em induced by the field}.  This is in
sharp contrast with the behaviour found both in conventional
superconductors \cite{Niobium} and in YBCO at higher temperatures
\cite{Palstra,Cohn,Ong}, where low fields (compared to H$_{c2}$) {\em
suppress} the conductivity.
\begin{center}
\begin{figure}
\resizebox{\linewidth}{!}{\includegraphics{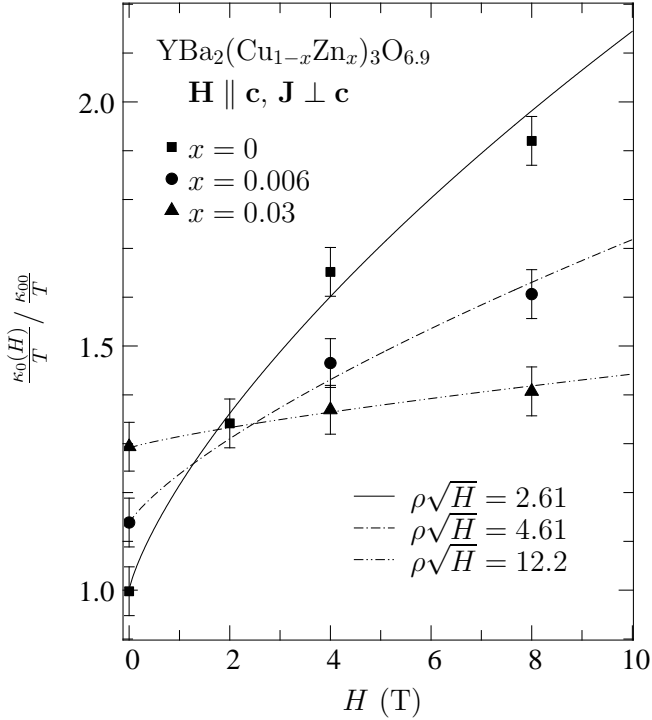}}
\vspace{0.5em}
\caption{Residual linear term $\kappa_0(H)/T$ normalized by
$\kappa_{00}/T$ as a function of applied field for pure (squares),
0.6\% Zn (circles) and 3\% Zn (triangles) samples.  Fits to Eq.(4) for
each crystal yield the values of $\rho$ shown (in units of
T$^{-1/2}$).}
\label{fig2}
\end{figure}
\end{center}

Volovik has shown how even at $T$=0 a small magnetic field can produce
quasiparticles near the nodes of a $d$-wave gap \cite{Volovik}.  In
the presence of a superfluid flow, the quasiparticle spectrum is
Doppler shifted to $E({\bf k},{\bf A}) = E({\bf k}) - \frac{e}{c} {\bf
v_k}\cdot {\bf A}$, where ${\bf v_k}$ is the normal state velocity and
{\bf A} is the vector potential \cite{LeeWen}.  In the case of
vortices, the magnitude of the Doppler shift may be characterized by
its average, $E_H$, obtained by integrating over a vortex-lattice unit
cell (of radius $R$): $E_H = \frac{e}{c} \langle {\bf v_k} \cdot {\bf
A} \rangle \sim \frac{1}{R^2} \int dr~r~v_F \cdot A \sim \frac{1}{R}
\Phi_0 v_F \sim \sqrt{H}$, using 2$R\simeq$ intervortex spacing $\sim
1/ \sqrt{H}$ and the fact that $\Phi_0 = \oint {\bf A}\cdot d{\bf
\ell}$, with $\Phi_0$ the flux quantum.  Volovik first derived the
$\sqrt{H}$ dependence and obtained the following field-induced density
of states (DOS) for a $d$-wave gap in the clean limit \cite{Volovik}:
$\delta N(0;H)/N_0\simeq \sqrt{8/\pi}~E_H/\Delta_0$, where $N_0$ is
the DOS at the Fermi level, $E_H=
a~\hbar~\sqrt{2/\pi}~v_F~\sqrt{H/\Phi_0}$ and $a$ is a vortex-lattice
parameter of order unity \cite{Kubert}.  The effect is sizable even at
low fields; in YBCO at 8~T, one expects $E_H\simeq$~40~K and $\delta
N(0;H)/N_0\simeq$~30\% (for $a$=1).

{\it Comparison with theory.}--- K\"ubert and Hirschfeld have
calculated the electronic thermal conductivity of a $d$-wave
superconductor in a field, neglecting vortex scattering
\cite{Kubert2}.  At $T$=0, they obtain for ${\bf H}~||~{\bf c}$ and
${\bf J} \perp {\bf c}$
\begin{equation}
\frac{\kappa(0;H)}{T}=\frac{\kappa_{0}}{T}\frac{\rho^2}
{\rho\sqrt{1+\rho^2}-\sinh^{-1}{\rho}}
\end{equation}
where $\rho$ is essentially the ratio of the two relevant energy
scales, $\gamma$ and $E_H$; in the dirty limit where $E_H < \gamma$,
$\rho = \sqrt{6/\pi}~\gamma/E_H \propto 1/\sqrt{H}$.  A one-parameter
fit of the data to Eq.(4) for each crystal is shown in Fig. 2. The
first point to note is that the sub-linear dependence on field is well
reproduced.  Perhaps more important is the fact that the magnitude of
the response in all three cases is very much as expected.  Indeed, the
fits yield the following values for $\rho$, evaluated at 8 T: 0.92,
1.63, and 4.32 for $x =$~0, 0.006, and 0.03, respectively.  These
correspond to a ratio $\gamma/E_H$ = 0.67, 1.18 and 3.12, which shows
that none of the crystals is in the clean limit over the field range
investigated.  Treating impurity scattering in the unitarity limit,
the scattering rate becomes: $\hbar \Gamma / k_B T_{c0}$~=~0.02, 0.07,
and 0.5, respectively (taking $\Delta_0 = 2.14k_B T_{c0}$), given that
$E_H\simeq$~20~K at 8~T (assuming $v_F$ = 1$\times$10$^7$ cm/s and
$a$=1/2).  These values are in remarkable agreement with those given
above as independent estimates based on the residual resistivity.

The same theoretical treatment has been applied to the specific heat.
K\"ubert and Hirschfeld have obtained $N(0;H)/N_0 =
\delta\gamma^*/\gamma_n = \sqrt{8/\pi}a\sqrt{H/H_{c2}}$, where
$\delta\gamma^*$ is the increase in specific heat due to an applied
field and $\gamma_n$ is the normal state specific heat at the Fermi
level \cite{Kubert}.  We can re-formulate this expression in terms of
$a$ and $v_2$ such that $\delta\gamma^* = \beta\sqrt{H}$ where $\beta
= \frac{8k_B^2}{3\hbar} \frac{1}{\sqrt{\Phi_0}} \frac{a}{v_2}$.  So
now we have three equations for the three parameters $a$, $v_F$ and
$v_2$.  Using $\beta = 0.9$~mJ~K$^{-2}$~T$^{-1/2}$~mol$^{-1}$
\cite{Moler}, we find $v_2 = 2.2a\times$10$^6$~cm/s.  From
$\kappa_0(0;H)/T$, we have $av_F = 5\times$10$^6$~cm/s.  These two
expressions, together with Eq.(3), yield $a = 0.40$.  With this value
of $a$, $v_2 = 0.9\times$10$^6$~cm/s and $v_F = 1.2\times$10$^7$~cm/s.
All three values are very reasonable, so that the overall picture is
quantitatively convincing, lending strong support to the basic
mechanism of a Doppler shift by superflow around vortices.  More
generally, it validates the theory of transport in unconventional
superconductors; in particular, the assumption that in these
correlated electron systems scattering by impurities must be treated
in the unitarity limit of strong, resonant scattering appears to be
correct.

{\it Other measurements.}--- Let us now compare our results with other
$\kappa$ measurements.  Note first that previous low temperature
measurements in YBCO were done with the field in the basal plane where
the Volovik effect is expected to be much smaller.  The two studies
were in disagreement, with a 50\% increase in 8 T in one case
\cite{Bredl}, but no change detected in 6 T in the other \cite{Wand}.
At higher temperatures, the fact that $\kappa$ decreases with field
could come from a number of effects. Perhaps the most natural is
vortex scattering, as invoked in the case of Nb.  While this can apply
to both quasiparticles and phonons, the fact that the largely
electronic peak below $T_c$ is almost completely suppressed in 10~T
\cite{Palstra} suggests that the former suffer most of the impact.
Franz has shown that a disordered vortex lattice in a $d$-wave
superconductor can result in quasiparticles scattering off the
superflow \cite{Franz}.  At low temperature, however, we find no
indication of significant vortex scattering, in the sense that there
is good agreement between calculations neglecting vortex scattering
and our data on crystals for which the relative strength of impurity
and vortex scattering must differ markedly from one crystal to the
next.  Further work is needed to arrive at a coherent description of
quasiparticle transport in the vortex state which covers both regimes
of behaviour.

It is interesting to compare our results on YBCO with the
corresponding results on BSCCO, obtained recently by Aubin, Behnia and
co-workers \cite{Aubin}.  At temperatures below 0.7~K, an increase in
$\kappa/T$ with field is also found, again with a sublinear (roughly
$\sqrt{H}$) dependence.  What is striking is the {\it magnitude} of
the response.  The application of only 2~T increases $\kappa/T$ by
about 0.20 mW~K$^{-2}$~cm$^{-1}$. Now from our own measurements on
pure, optimally-doped single crystals of BSCCO, the (presumably
universal) residual linear term is $\kappa_0/T$ =
0.18$\pm$0.03~mW~K$^{-2}$~cm$^{-1}$ \cite{Lambert}.  This means that
in BSCCO a field of 2~T causes the quasiparticle conduction to {\it
double}, whereas it produces only a 35\% increase in our pure YBCO
crystal.  In reality, the field dependence in BSCCO is much more than
3 times stronger, since the impurity scattering rate in the crystal
used by Aubin and co-workers could be as much as 100 times larger.
Indeed, its residual resistivity is 130 $\mu\Omega$~cm \cite{Kamran},
compared with approximately 1 $\mu\Omega$~cm in our pure YBCO crystals
\cite{Taillefer}.  These considerations lead us to conclude that the
nature of defect scattering in these two (otherwise quite similar)
materials is significantly different.  Either the kind of defect found
in nominally pure crystals is different, or the impact that a given
defect (e.g. impurity) has on the surrounding electron fluid is
different.  It will be interesting to see whether the current
theoretical framework can account for the BSCCO data simply by
adjusting the impurity phase shift, moving it away from the unitarity
limit towards the Born limit.

In summary, we have measured the low temperature heat conduction by
quasiparticles in the CuO$_2$ planes of YBCO as a function of magnetic
field, for different impurity levels.  In all cases, the residual
linear term $\kappa_0/T$ is found to increase with field strength,
directly reflecting the additional population of extended
quasiparticle states.  The good agreement with calculations by
K\"ubert and Hirschfeld for a $d$-wave superconductor allows us to
draw the following conclusions: 1) the ``Volovik effect'' is fully
verified, and it is the dominant mechanism behind the field dependence
of transport in YBCO as $T\to$~0, for ${\bf H}~||~{\bf c}$; 2) vortex
scattering, invoked to explain the behaviour at intermediate
temperatures, is seemingly absent at low temperature; 3) the
widespread assumption that impurities (or defects) can be treated as
unitary scatterers in correlated electron systems appears to be
verified in YBCO; 4) the nature of impurity scattering differs between
BSCCO and YBCO.

We thank K. Behnia, P. Hirschfeld and P. A. Lee for insightful
discussions, and B. Lussier for his help during the early stages.
This work was supported by the Canadian Institute for Advanced
Research, and funded by NSERC of Canada and FCAR of Qu\'ebec.  MC
acknowledges support from the Carl Reinhardt Foundation.

\vspace{0.5em}\noindent
$^\dagger$ Current address: Department of Physics, University of
Toronto, Toronto, Canada M5S 1A7.

\end{document}